# Pressure-induced stabilization of carbonic acid and other compounds in the C-H-O system

*Gabriele Saleh[a,*], Artem R. Oganov[a,b,c,d]*

a Moscow Institute of Physics and Technology, 9 Institutsky Per., Dolgoprudny, Moscow Region, 141700, Russia.

b Skolkovo Institute of Science and Technology, Skolkovo Innovation Center, 3 Nobel St., Moscow 143026, Russia.

c Department of Geosciences and Department of Physics and Astronomy, Stony Brook University, Stony Brook, New York 11794-2100, USA

d International Center for Materials Discovery, Northwestern Polytechnical University, Xi'an, 710072, China

Correspondence and requests for materials should be addressed to G. S.
(email: gabrielesaleh@outlook.com)

ABSTRACT

The physicochemical behavior of elements and compounds is heavily altered by high pressure. The occurrence of pressure-induced reactions and phase transitions can be revealed by crystal structure prediction approaches. In this work, we explore the C-H-O phase diagram up to 400 GPa exploiting an evolutionary algorithm for crystal structure predictions along with *ab initio* calculations. Besides uncovering new stable polymorphs of high-pressure elements and known molecules, we predicted the formation of new compounds. A $2CH_4:3H_2$ inclusion compound forms at low pressure and remains stable up to 215 GPa. Carbonic acid ($H_2CO_3$), highly unstable at ambient conditions, was predicted to form exothermically at mild pressure (about 1 GPa). As pressure rises, it polymerizes and, above 300 GPa, reacts with water to form orthocarbonic acid ($H_4CO_4$). This unexpected high-pressure chemistry is rationalized by analyzing charge density and electron localization function distributions, and implications for general chemistry and planetary science are also discussed.

The dramatic influence of high pressure (tens or hundreds of GPa) on reactivity is experimentally well established[1]. Crystal structure prediction approaches have become very effective in correctly anticipating experimental outcomes[2]. Among the numerous applications, these techniques have been widely exploited to predict high-pressure reactions[3,4,5]. Composition-pressure phase diagrams can be built by comparing the free energy (which at T = 0 K reduces to enthalpy) of the most stable structures of elements and compounds at various pressures, in order to single out the thermodynamically stable compositions. The latter are defined as those compounds for which no exothermic decomposition reactions exist. Often, binary phase diagrams are targeted, for the high number of possible stoichiometries in a ternary phase diagram makes its *ab initio* exploration very computationally demanding.

The C-H-O ternary phase diagram at high pressure is of paramount interest for planetary science. $H_2O$ and $CH_4$, not necessarily in their intact molecular forms, are the major constituents of giant planets such as Neptune and Uranus, where pressure can reach values of hundreds of GPa[6]. Moreover, large icy satellites (*e.g.* Ganymede, Callisto and Titan)[7] and comets[8] all contain water ice, mixed with a number of volatiles such as $CH_4$ and $CO_2$, experiencing pressures up to a few GPa.

Some information on the C-H-O phase diagram is known. At ambient pressure, all compounds but water, carbon dioxide and methane are thermodynamically unstable, *i.e.* their decomposition into a mixture of other molecules is invariably exothermic. Whereas the former two compounds are expected to survive even in the terapascal regime[9,10], above 155 GPa methane was predicted to decompose into mixtures of hydrogen and heavier hydrocarbons, namely ethane, butane and polyethylene[11]. On further compression, all alkanes disproportionate into diamond and hydrogen[11,12]. The low-pressure part of the C-H-O phase diagram is particularly intricate for it incorporates a wide variety of inclusion compounds. Above 4 GPa, $CH_4$ and $H_2$ combine in various proportions to form molecular co-crystals[13]. When water is exposed to gases at low temperature and moderate pressures (typically 6-15MPa)[14], gas hydrates may form. The latter are widely known compounds made by a framework of hydrogen-bonded water molecules encapsulating guest molecules. Common guests within the C-H-O composition are oxygen, hydrogen, carbon dioxide and small hydrocarbons. All these inclusion compounds are known to decompose when pressure rises beyond certain limits, typically a few GPa. However, some exceptions were reported[15], *e.g.* the $H_2O:2H_2$ complex was predicted to be stable "up to at least 120 GPa"[16]. Within inclusion compounds, host and guest retain their molecular identities and interact through weak van der Waals forces. In fact, no thermodynamically stable molecules formed solely by C, H and O altogether are known to date.

In this work, the C-H-O phase diagram is explored in the pressure range 10-400 GPa by means of the powerful variable-composition evolutionary algorithm USPEX[17] coupled with periodic density-functional calculations. Thorough, unbiased searches were performed sampling all possible C-H-O compositions, and a total of more than 125000 structures, generated by the evolutionary method, were relaxed to the closest minimum-enthalpy configuration. Chemical bonding in the resulting compounds was investigated by analyzing the real-space distribution of their charge density, in the framework of the Quantum Theory of Atoms in Molecules (QTAIM[18]), and of their electron localization function (ELF[19]). According to QTAIM, all the properties of a system can be partitioned into atomic contributions. Moreover, chemical bonds are mirrored in the charge density distribution as special saddle points, called 'bond critical points' (bcps). The evaluation of certain scalar properties at bcps provides information about the chemical bond type. ELF is a simple measure of electron localization. Maxima appear in those regions associated with chemical bonds, lone pairs and atomic cores, and electronic populations can be assigned to these chemical entities by integrating charge density within the basins corresponding to those maxima.

A schematic representation of the C-H-O pressure-composition phase diagram is reported in Fig. 1a-b. Our calculations recovered all the stable compounds known from previous crystal structure prediction works[9,10,11,16,20,21,22]. As for the crystal structures, we obtained either the same as the previously reported, or enthalpically (typically within 1 meV/atom) and structurally similar ones. A detailed comparison is reported in Supplementary Information (Section S2), while some cases are worth being mentioned here. In both $H_2O:H_2$ and $H_2O:2H_2$ hydrates, the water framework is based on the structure of ice Ic[16]. The structures we obtained for these two compounds are similar to the previously reported ones and differ only by the orientation of $H_2$ guest molecules or by a donor-acceptor switch in the O-H···O interaction. It is noteworthy that above 100 GPa, our $H_2O:2H_2$ crystal structure (Fig. 2a) and the reported one have different $H_2$ orientation but their enthalpy is indistinguishable (ΔE≤ 0.5 meV/atom), suggesting that the guest molecules are rotationally disordered even at high pressure. For this compound, we determined for the first time its decomposition pressure: 153 GPa. We discovered a structure of ethane (Fig. 2b) that is more enthalpically favorable than the previously reported one. This reduces the previous estimate[11] of the pressure of the decomposition reaction $2CH_4 \rightarrow C_2H_6 + H_2$ from 200 GPa down to 121 GPa. For water ice above 200 GPa, we obtained the Pbcm structure already proposed in previous *ab initio* studies[23,24] (Fig. 2d). Concerning oxygen, our high-pressure calculations resulted in the previously unknown hexagonal structure reported in Fig. 2e ($P6_3/mmc$ space group), stable above 375 GPa. Similarly to its lower-pressure ξ-phase (space group C2/m)[20], this $P6_3/mmc$ oxygen is a metallic molecular crystal (Fig. S2). In the lower-pressure ε- and ξ-phases, the intermolecular distances within the *ab* crystallographic plane have different values, ε-phase even having exotic $(O_2)_4$ clusters[20]. In the $P6_3/mmc$ structure, instead, they become equal thereby conferring to the crystal its hexagonal symmetry.

New compounds were predicted to form. A $2CH_4:3H_2$ (Fig. 2c) clathrate was found to be stable from < 10 GPa up to 215 GPa. Its crystal structure does not vary throughout its pressure range of stability. Curiously, the topology of the host framework is the same as in the $H_2O:H_2$ gas hydrates discussed above (Fig. S15), despite the absence of hydrogen bonds (HBs) in $CH_4:H_2$ inclusion compounds. Concerning the possible formation of molecules containing C, H and O, our results indicate that carbonic acid becomes thermodynamically stable above 0.95 GPa (effects of zero-point energy correction and temperature are discussed below). This was quite unexpected as this molecule is highly unstable at ambient conditions[25]. Its synthesis requires the use of high-energy radiation or strong acids and the resulting compound can only be isolated under high vacuum or in argon matrix and at very low temperatures[26,27]. Indeed, the decomposition of carbonic acid into water and carbon dioxide is highly exothermic, although in absence of water it is hampered by a high kinetic barrier[28]. The low-pressure structure (Pnma space group, Fig. 3a) is composed of chains made by hydrogen-bonded molecules. The latter adopt an almost flat conformation in which hydrogens are in *cis* position with respect to the C=O bond. This type of HBs array was shown to be the most energetically favorable at ambient conditions, 'cis-cis' being the most stable conformation of the isolated molecule[27]. The crystal structure of carbonic acid is significantly denser than water ice (2.147 *vs* 1.560 g/cm$^3$ at 1 GPa). This fact has important implications for planetary science (*vide infra*). As pressure rises, carbonic acid polymerizes (p> 44 GPa), forming the structure shown in Fig. 3b. The -CO- backbone of each polymer is parallel to the *c* crystallographic axis, while hydroxyl groups form HBs joining the adjacent polymers along the *b* direction. This type of structure is stable up to the highest investigated pressure (400 GPa), although the HB network rearranges above 240 GPa, leaving the CO backbone and the crystallographic space group unchanged. In the higher-pressure conformation (Fig. 3c), each polymer is bonded to the nearest neighbors along the two *ab* diagonals. This implies that, differently from the lower-pressure Cmc2$_1$ structure, at high pressure the HB network joins all the polymers together. Finally, at 314 GPa, we detected an exothermic reaction between carbonic acid and water to form orthocarbonic acid (H$_4$CO$_4$, Fig. 3d).

Besides being thermodynamically stable, the newly discovered compounds were ascertained to be dynamically stable by the absence of imaginary frequencies in their phonon dispersion curves (section S6).

The intricate chemistry described above can be rationalized by taking a closer look at the crystal structures and by analyzing the QTAIM properties and ELF distribution. First we note that, as molecular crystal Pnma-H$_2$CO$_3$ is compressed, the C-O hydroxyl bonds (C=O carbonyl bonds) shorten (elongate) and their bcps ellipticity increases (decreases), as shown in Table S2. This indicates a progressive delocalization of the π orbital over the two formally single C-O1 bonds, eventually making them shorter than the formally double C=O2 bond. This behavior can be explained with the chemical scheme of Fig. 3e. As pressure rises, the weight of the resonance forms II and III increases. ELF distribution supports this hypothesis: a fourth maximum appears around O2 above 40 GPa (Fig. 4c-f), indicating a progressive shift from sp$^2$ to sp$^3$ hybridization, and at 100 GPa 5.8 out of 7.6 valence electrons are contained in the lone pairs basins (Table 1). This trend can be seen as a pressure-induced destabilization of the double bond, which is likely to be an important factor contributing to the carbonic acid polymerization. The latter is indeed associated with the breaking of the π bond and the concurrent formation of a new C-O σ bond, in a way which is reminiscent of the high-pressure behavior of CO$_2$[9]. Accordingly, along the Pnma→Cmc2$_1$ phase transition, C-O bonds elongate and their elipticity at bcp decreases (Table 1). HBs undergo important changes, too. Differently from the Pnma phase, in the Cmc2$_1$ structure O-H and H⋯O interactions display similar bond lengths, degree of covalency (as measured by the kinetic energy density per electron[29]) and electron density at bcp (Table 1). Moreover, in passing from Pnma to Cmc2$_1$ phases, a significant decrease (increase) in the population of the ELF basins corresponding to acceptor lone pairs (O-H bonds) takes place. Such a charge transfer from the acceptor lone pair to the donor-H fragment generally occurs in passing from weak to strong HBs[30]. Thus, along the Pnma→Cmc2$_1$ phase transition, HBs strengthen and become noticeably more symmetric. The symmetrization of HBs becomes more effective as pressure rises, and in the Cmc2$_1$ phase at 400 GPa O-H and H⋯O interactions are nearly indistinguishable (Table 1). The different nature of HBs in the two phases of carbonic acid is even more evident in the ELF distribution (Fig. 4a-b), which for Cmc2$_1$ is more akin to that of symmetric HBs observed in the high-pressure phases of ice (ice X and ice Pbcm, Fig. S7). Noteworthy is the appearance of an ELF maximum on H atoms for all the investigated high-pressure HBs, a feature observed at ambient pressure only in 3c-2e bonds (e.g. in F-H-F$^-$ and H$_2$O-H-OH$_2^+$)[31].

The reaction H$_2$CO$_3$+H$_2$O → H$_4$CO$_4$ becomes exothermic at high pressure due to the volume reduction (i.e. is driven by the pV term in the free energy) and is homodesmic: the total number of C-O, O-H and H⋯O bonds remains constant, and so does the total number of oxygen's free lone pairs (i.e. lone pairs not involved in any HB). Therefore, insights into the stabilization mechanism of orthocarbonic acid can be gained by studying how population and charge of the various types of ELF basins vary along its formation reaction (Fig. 5). The valence basins of water and ether oxygens undergo the greatest changes. Overall, in passing from reactants to orthocarbonic acid, the basins of free lone pairs sizably increase both their volume and electron population, while a small expansion and a roughly constant population are observed for C-O bonds (Fig. 5, inset). The remaining basins, i.e. those of O-H⋯O interactions, lose a corresponding amount of charge. However, their major shrinking overcompensates the expansion observed for C-O and free lone pairs, thereby resulting in a net volume reduction upon H$_4$CO$_4$ formation. These results indicate that the increase in charge concentration of free lone pairs, which are not present in water due to the ice rule, and their interplay with HBs, have a crucial role in the high-pressure stability of orthocarbonic acid. This hypothesis is supported by the product *vs* reactants comparison of QTAIM atomic basins (Table S3): water oxygen undergoes a notable shrinking upon formation of $H_4CO_4$, while the volume changes of other atoms are minor. We observe similar ELF and QTAIM trends in the other detected polymorph of $H_4CO_4$ (Section S8), which is structurally different from and less stable than orthocarbonic acid, but is enthalpically favorite over the $H_2O + H_2CO_3$ mixture above 395 GPa.

We now move to discuss the implications of our results for general chemistry and planetary science. The long-standing view that inclusion compounds systematically decompose at low pressures of a few GPa has been refuted in light of a number of counter-examples discovered during the last decades. However, only hydrates were known to persist above 50 GPa[15]. Our $2CH_4:3H_2$ compound not only sets a new upper limit for the stability of inclusion compounds, but also introduces a qualitative shift of views, for it broadens the classes of inclusion compounds stable at very high pressures. In fact, several $nCH_4:mH_2$ crystals were experimentally found to be stable up to 8 GPa, and $CH_4:2H_2$ was compressed up to 30 GPa without showing any sign of decomposition[13]. The discrepancy in the stoichiometry between such compound and the one presented here might be traced back to a number of causes, for example the possible metastability of the experimentally detected phase above 10 GPa. Remarkably, the existence of $2CH_4:3H_2$ affects the high-pressure chemistry of methane: (1) it lowers the decomposition pressure of pure methane crystals down to 93 GPa, much lower than previous estimates[11] (Fig. 1a), and (2) allows methane molecules to survive up to 221 GPa (Fig. 1b), *i.e.* at a higher pressure higher than previously reported.

Concerning carbonic acid, its discovered stability at moderate pressure opens important possibilities for new synthetic pathways and new ways of stabilizing this evanescent molecule. However, it must be pointed out that the direct comparison between the calculated and experimental formation pressures is complicated by two problems: the influence of lattice vibrations (temperature effects and zero-point energy correction) and the approximations made within the DFT approach. We have tackled these two issues by evaluating how the formation pressure of $H_2CO_3$ is affected by phonons (within the harmonic approximation) and by the use of different computational settings (including changes in the basis set and exchange-correlation functional). The pressure required for stabilizing carbonic acid shifts to 1.45 GPa when zero-point vibrational energy is accounted for, and then weakly increases with temperature (Fig. 1c-d). All the tested computational approaches reproduced the pressure-induced stabilization of carbonic acid, although the formation pressure showed some variations (Table S5). Overall, a realistic estimation for the formation pressure of carbonic acid would be the range 0.6-1.6 (0.75-1.75) GPa at 100 (300) K. Similar pressures occur on the bed of water oceans of icy satellites[7]. There, both water ice and carbon dioxide are present, hence carbonic acid is likely to form. Moreover, its high density implies that, once formed, carbonic acid will sink to the bottom of ice layers, just above the rocky cores, thereby experiencing an even greater pressure. In such a scenario, carbonic acid insulates water ice from the core. This would modify the chemical composition models for these celestial bodies, which now include possible reactions between water and rocky compounds such as ferromagnesiansilicates[7].

In conclusion, we have carried out a thorough DFT investigation on the C-H-O phase diagram up to 400 GPa. The most stable structures of each stoichiometry at various pressures were obtained by the powerful evolutionary algorithm USPEX. Besides those already known, our calculations predicted the formation of several new compounds. An inclusion compound, $2CH_4:3H_2$, was found to be stable up to the unprecedented pressure of 215 GPa. Carbonic acid was predicted to be stable above 1 GPa, to remain stable throughout the investigated pressure range and to polymerize above 44 GPa. At 314 GPa it reacts with water to form orthocarbonic acid, $H_4CO_4$. A thorough chemical bonding analysis was performed, which provided important insights such as the pressure-induced stabilization of resonance forms of molecular carbonic acid and the paramount role of lone pairs in the stabilization of orthocarbonic acid. This novel chemistry can have major implications for planetary science.

METHODS

The variable-composition evolutionary approach implemented in USPEX, scans the structural and chemical spaces and seeks the thermodynamically stable compounds. The first generation of structures was mostly generated randomly (stable structures obtained from previous runs were also added). The successive generations were produced by both generating new random structures and by applying variation operators (as described in ref.32) to the most stable compounds (65% of the total population). Details on each USPEX run carried out in this work can be found in Table S1. Exhaustive description of the method[32,33] and studies where the approach reliability was confirmed by comparison with experiments[3,11,20,34] are reported elsewhere. The VASP code[35] was used for structure relaxations and total enthalpy calculations. We adopted the Perdew-Burke-Ernzerhof (PBE) functional[36] in the framework of the all-electron projector augmented wave (PAW) method[37], with 'hard' PAW potentials, plane wave kinetic energy cutoff of 850 eV and a uniform $\Gamma$-centered grid with $2\pi*0.056$ Å$^{-1}$ spacing for reciprocal space sampling. Below 10 GPa, where dispersion interactions play an important role, we employed for VASP calculations a van der Waals functional (optB88-vdw[38]). At these pressures, a slightly looser reciprocal space sampling of $2\pi*0.064$ Å$^{-1}$ was adopted. Note that at each pressure the stability of each compound was ascertained by considering the most stable form of the reactants, including structures from literature. To perform the chemical bonding analysis, we carried out single-point calculations (at the geometry obtained from VASP) with the CRYSTAL14 code[39]. Within the latter, crystal orbitals are described in terms of atom-centered functions. The employed basis set was of 'triple-$\xi$ plus polarization' quality, whose functions were optimized for solid-state calculations[40]. In order to better describe intermolecular interactions, we augmented the basis set with d orbitals on H atoms taken from ref. 41. To make the basis set apt for high-pressure calculations, the exponents of the radial part of the outermost functions of each shell were contracted by a factor 1.44 and 1.69 for calculations below and above 200 GPa, respectively. The bond critical point analysis and the determination of atomic basins within the QTAIM framework was done using the TOPOND code, now implemented within CRYSTAL14. The latter was also exploited to evaluate grid files of scalar properties (such as charge density and ELF), subsequently converted in the standard 'cube' format using the NCImilano code[42]. The integration of quantities within ELF basins, not implemented in TOPOND, was performed with the critic2 code[43], and in particular the grid-based Yu-Trinkle algorithm[44] was exploited. The input grids were 400x400x400, spanning the whole unit cell. Phonon dispersion curves, and phonon contribution to the free energy of formation were calculated by means of the finite displacement method implemented in the PHONOPY code[45]. Further details concerning phonon calculations are reported in Section S6. Images of structures and 2D maps/isosurfaces were produced with Diamond[46] and VESTA,[47] respectively.

ACKNOWELEDGMENTS

This work was supported by the grant of the Government of the Russian Federation (No. 14.A12.31.0003). We acknowledge the use of supercomputer "Lobavchevsky" of the State University of Nizhny Novgorod (Russian Federation) and of the Rurik supercomputer of our laboratory at MIPT.

AUTHORS CONTRIBUTIONS

G.S. performed and analyzed the calculations. Both authors designed the research and wrote the manuscript.

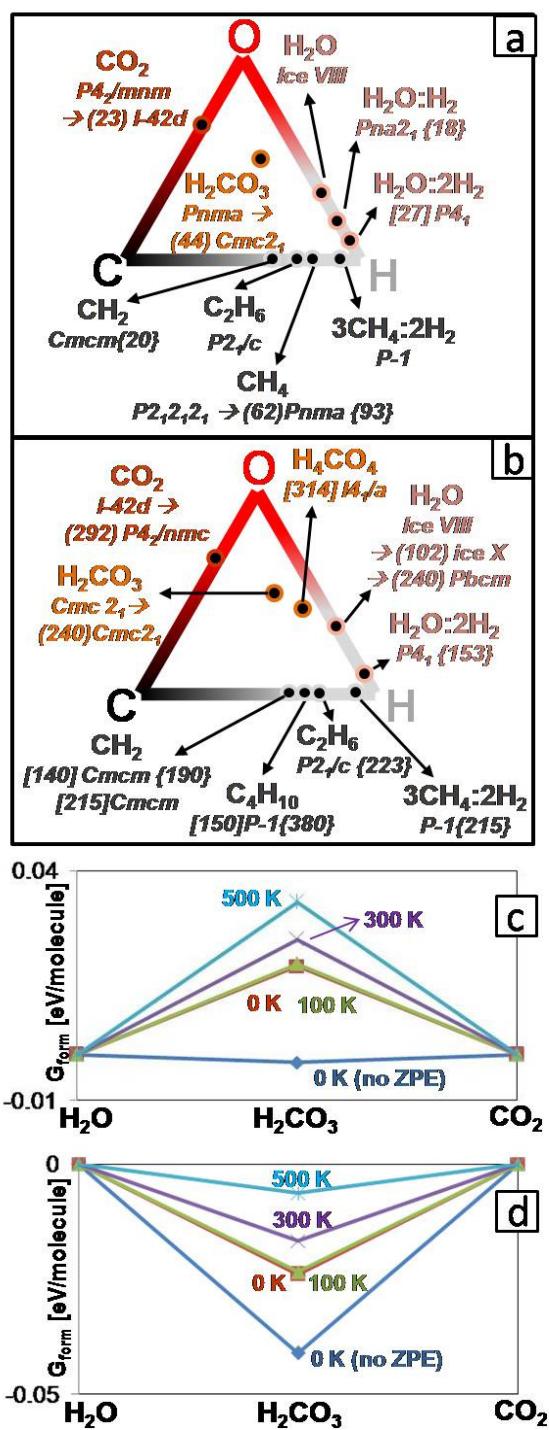

**Figure 1. Summary of stable phases in the C-H-O phase diagram and convex hull for the $CO_2$-$H_2O$ system.** Stable phases in the C-H-O system are reported for the ranges 10-100 GPa (a) and 100-400 GPa (b). For each compound, we report the space group(s) and the predicted phase transition pressures (in GPa), the latter in round brackets. Numbers in square (curly) brackets indicate the formation (decomposition) pressures. In (a), the formation pressure is not reported for those compounds already stable at 10 GPa. For elements, a detailed description of the known phases and the comparison with the ones obtained by our calculations are reported in Sect. S2. A complete representation of the C-H-O phase diagram at various pressures, including Gibbs triangles, is reported in Fig. S8. We also report the convex hulls at various temperatures for the $H_2O$-$CO_2$ system at 1 GPa (c) and 2 GPa (d). 'no ZPE' represent the values obtained neglecting the zero-point (vibrational) energy correction.

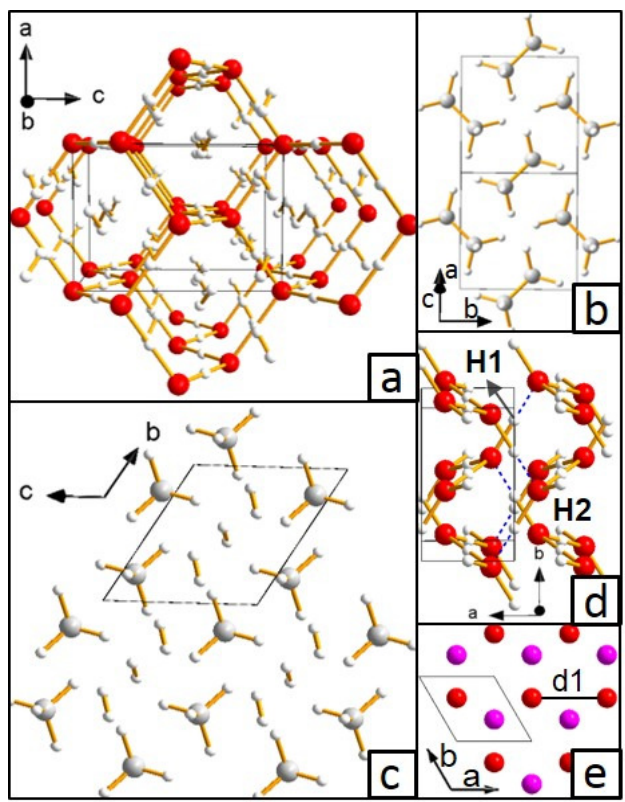

**Figure 2. Novel crystal structures in the C-H-O phase diagram.** (a) $H_2O$:$2H_2$, $I4_1a$ space group. (b) ethane, space group $P2_1/c$. (c) $2CH_4$:$3H_2$ inclusion compound, $P-1$ space group. This structure is also reported in Fig. S15, where the host framework topology is highlighted. (d) Pbcm-$H_2O$. There are 3 symmetry-independent atoms: 2 H (labeled in the picture) and 1 O (e) high pressure phase of oxygen, $P6_3/mmc$ space group. $O_2$ molecules are oriented along the *c* axis. Atoms in background are colored in violet. The distances between nearest neighbor molecules (d1 in the picture) are 1.878 Å at 400 GPa. In this and other pictures, C, H, and O atoms are colored in grey, white and red, respectively. Bonds are represented as yellow sticks while H···O interactions of asymmetric hydrogen bonds are indicated as dotted blue lines.

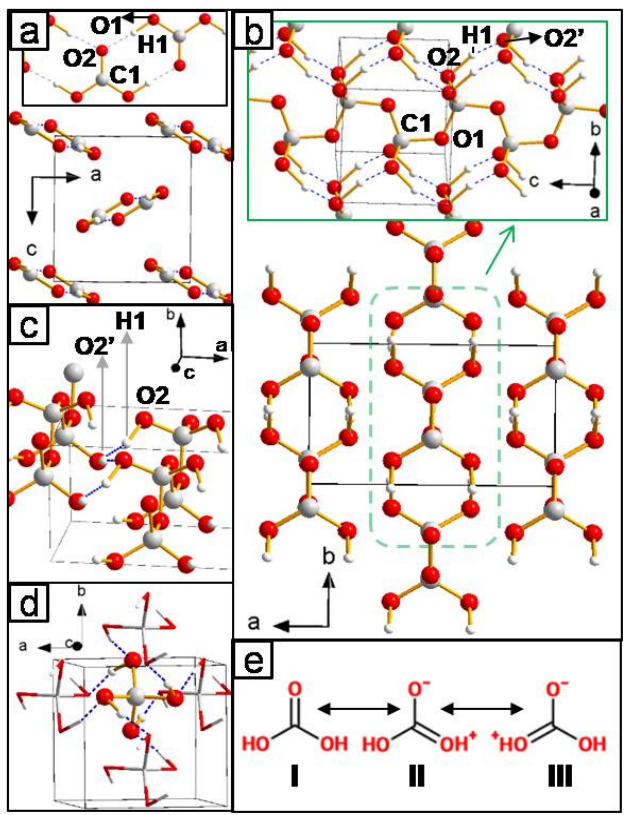

**Figure 3. Crystal structures and chemical schemes for carbonic acid and H$_4$CO$_4$.** (a) Pnma-H$_2$CO$_3$. Chains of hydrogen-bonded molecules extend along the *b* axis, and they are shown in the inset. The resonance forms discussed in the main text are shown in (e). (b) Cmc2$_1$-H$_2$CO$_3$, low-pressure form. Polymers extend along the *c* axis, and they are shown in the inset. (c) hydrogen bonds in the high-pressure form of Cmc2$_1$-H$_2$CO$_3$ (d) I4$_1$/a-H$_4$CO$_4$. We show a single molecule (represented in 'ball-and-stick' model), along with all its hydrogen-bonded neighbors ('stick model'). Additional representation of the crystal packing are reported in Fig. S16. For all the structures having more than one symmetry-independent atom of each type, the labels adopted in Fig. 4-5 and Table 1 are shown.

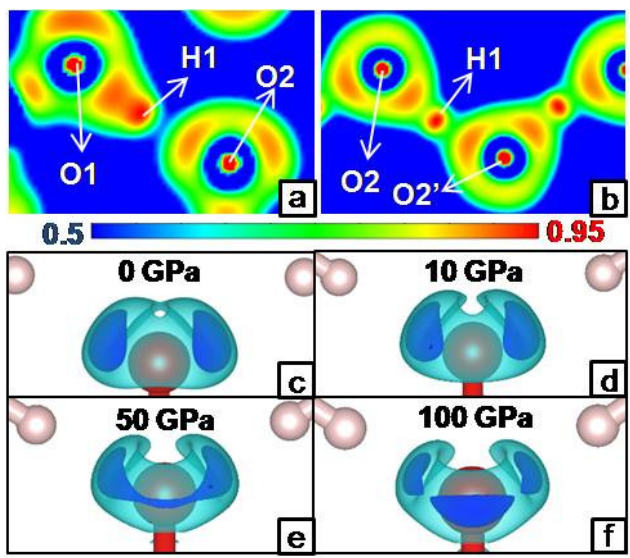

**Figure 4. ELF distribution in carbonic acid.** The ELF distribution at 100 GPa in the plane containing the O-H⋯O interaction is shown for Pnma-$H_2CO_3$ (a) and Cmc$2_1$-$H_2CO_3$ (b). The color scale is reported below the pictures. (c)-(f) ELF isosurfaces showing the lone pairs of the O2 atom in Pnma-$H_2CO_3$ at various pressures (isosurface relative to C-O bond omitted for clarity). The adopted isovalues are 0.850 (light blue) and 0.875 (dark blue). Further 3D and 2D representations of ELF distribution in $H_2CO_3$, $H_4CO_4$ and $H_2O$ can be found in Figs. S6-S7.

**Table 1. ELF and QTAIM properties of symmetry-independent bonds for selected compounds.**

| bond, A-B | $d_{AB}$ [Å] | $\rho_{BCP}$ [e/$a_0^3$][a] | $G_{BCP}/\rho_{BCP}$ [$E_h$/e][b] | ELF A-B: V[$a_0^3$];n[c] | ELF l.p.: V [$a_0^3$], n[d] |
|---|---|---|---|---|---|
| Cmc$2_1$-$H_2CO_3$(100GPa) | | | | | |
| O1-C1 | 1.363 | 0.31 {0.05} | 0.79 | 7.94;1.48 | 19.3;2.40 |
| C1-O1 | 1.336 | 0.32 {0.03} | 0.90 | 8.01;1.52 | - |
| C1-O2 | 1.325 | 0.34 {0.03} | 0.85 | 8.36;1.57 | 21.2;2.55 |
| O2-H1 | 1.134 | 0.22 | 0.51 | 11.8;1.61 | *3.27;0.28* |
| H1⋯O2 | 1.165 | 0.20 | 0.58 | 13.9;1.84 | - |
| Pnma-$H_2CO_3$(100GPa) | | | | | |
| C1-O2 | 1.260 | 0.38 {0.11} | 1.31 | 10.8;1.85 | 16.1;1.82[e] |
| C1-O1 | 1.234 | 0.40 {0.16} | 1.48 | 12.6;2.04 | 16.4;1.96 |
| O1-H1 | 1.053 | 0.28 | 0.39 | 11.6;1.59 | *4.96;0.38* |
| H1⋯O2 | 1.261 | 0.15 | 0.69 | 15.9;1.98 | - |
| Cmc$2_1$-$H_2CO_3$(400GPa) | | | | | |
| O1-C1 | 1.286 | 0.36 | 1.23 | 6.50;1.53 | 12.83;2.34 |
| C1-O1 | 1.277 | 0.37 | 1.26 | 7.12;1.58 | - |
| C1-O2 | 1.236 | 0.41 | 1.39 | 7.49;1.70 | 14.2;2.56 |
| O2-H1 | 1.063 | 0.27 | 0.65 | 8.61;1.66 | *2.42;0.30* |
| H1⋯O2 | 1.066 | 0.27 | 0.66 | 8.33;1.63 | - |
| I4$_1$/a-$H_4CO_4$ (400GPa) | | | | | |
| C-O | 1.268 | 0.38 | 1.24 | 7.41;1.63 | 13.9;2.56 |
| O-H | 1.058 | 0.28 | 0.61 | 8.51;1.67 | *2.64;0.32* |
| H⋯O | 1.096 | 0.25 | 0.69 | 8.46;1.66 | - |
| Pbcm-$H_2O$ (400 GPa) | | | | | |
| O-H2 | 1.044 | 0.29 | 0.68 | 10.1;1.87 | *2.31;0.30* |
| O-H1 | 1.033 | 0.30 | 0.66 | 10.1;1.80 | *2.30;0.30* |
| H1⋯O | 1.054 | 0.28 | 0.70 | 9.26;1.75 | - |

(a) electron density at the bond critical point. For C-O bonds at 100 GPa, we report in curly brackets their ellipticity. The latter is defined as $\varepsilon(r)=[\lambda_1(r) / \lambda_2(r)]-1$, where $\lambda_n$ is the n-th lowest eigenvalue of the electron density Hessian matrix. Ellipticity measures the deviation of electron density distribution along the internuclear axis from the cylindrical symmetry characteristic of σ bonds. (b) kinetic energy density per electron. (c) volume and electron population (in this order) of the disynaptic ELF basins joining atoms A and B. We identify O-H⋯O hydrogen bonds as those atom triads for which the OHO angle and the H⋯O distance are greater than 130° and lower than 2 Å, respectively. The ELF basin associated to H⋯O corresponds to the HB acceptor lone pair. (d) in this column we report the volume and electron population (in this order) of the ELF basins corresponding either to the 'free lone pairs' of the oxygen involved in the A-B bond, or to the basin centered on the H nucleus. In the latter case, the numbers are in italic. We define as 'free lone pairs' the lone pairs of oxygen which are not acceptors of hydrogen bonds. For O atoms not acting as hydrogen bond acceptor (hence formally having 2 free lone pairs), we report the total lone pair population divided by two. e) note that, at 100 GPa, Pnma-$H_2CO_3$ displays 3 lone pairs on O2, one of which does not act as hydrogen bond acceptor.

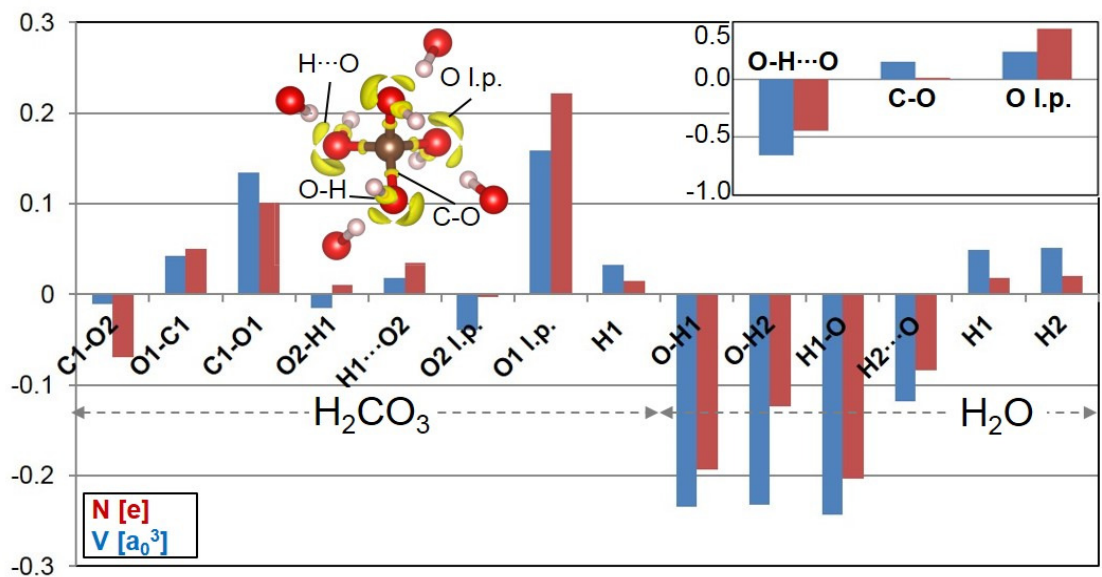

**Figure 5. Variation of properties of valence ELF basins along the reaction $H_2O + H_2CO_3 \rightarrow H_4CO_4$.** The histogram shows the difference in the electron population (red) and volume (blue) of ELF basins between orthocarbonic acid and the reactants at 400 GPa. For each symmetry-independent basin of the reactant, we subtract its volume/charge from that of the basin of the same type (C-O, O-H and H⋯O bonds, H atoms and free lone pairs of oxygen, the latter being labeled as 'l.p.' in the plot) of orthocarbonic acid, the latter having only one symmetry-independent basin of each type. A representative ELF isosurface plot of one $H_4CO_4$ molecule in the crystal is also shown (isovalue = 0.86), along with the O-H fragments of neighboring molecules which act as hydrogen-bond donor. The isosurfaces corresponding to the each type of valence basin are labeled (except for the one on the H position, which is hidden inside the sphere representing the H atom). The up-right inset shows the total variation of various types of basins (O-H⋯O is the sum of O-H, H, and H⋯O basins) along the reaction.